\documentclass{aa}

\usepackage{graphics}

\begin{document}

   \thesaurus{		  % A&A 
              (03.20.9;   % Telescopes
               03.20.1;   % Techniques: interferometric
               03.09.2;   % Instrumentation: interferometers
               09.07.1)}  % ISM: general

   \title{The Synthesis Telescope at the Dominion Radio Astrophysical
Observatory}

   \author{T.L. Landecker \inst{1} \and P.E. Dewdney \inst{1} \and
T.A. Burgess \inst{1} A.D. Gray \inst{1} \and L.A. Higgs \inst{1} 
\and A.P. Hoffmann \inst{1} \and G.J. Hovey \inst{1} \and D.R. Karpa
\inst{1,} \inst{2,} \inst{3} \and J.D. Lacey \inst{1} \and N. Prowse \inst{1,}
\inst{4} \and C.R. Purton \inst{1} \and R.S. Roger \inst{1} \and
A.G. Willis \inst{1} \and W. Wyslouzil \inst{1} \and D. Routledge
\inst{2} \and J.F. Vaneldik \inst{2}}

   \offprints{T.L. Landecker}

   \institute{National Research Council Canada, Herzberg Institute of
Astrophysics, Dominion Radio Astrophysical Observatory, Penticton, B.C.
Canada  V2A 6K3\\ email: tom.landecker@hia.nrc.ca
   \and
Electrical and Computer Engineering Department, University of Alberta,
Edmonton, Alberta, Canada T6G 2G7
   \and
Present address: Rockwell Collins, Cedar Rapids, Iowa, USA 52498
   \and
Present address: NetFacet Computing Inc., 492 Fraser Ave., Ottawa, Ontario,
K2A 2R2, Canada}

   \date{Received April 4, 2000; accepted June 23, 2000}

   \authorrunning{Landecker et~al.}

   \titlerunning{The DRAO Synthesis Telescope}

   \maketitle

   \begin{abstract}

We describe an aperture synthesis radio telescope optimized for
studies of the Galactic interstellar medium (ISM), providing the
ability to image extended structures with high angular resolution over
wide fields.  The telescope produces images of atomic hydrogen
emission using the 21-cm \ion{H}{i} spectral line, and,
simultaneously, continuum emission in two bands centred at 1420\,MHz
and 408\,MHz, including linearly polarized emission at 1420 MHz, with
synthesized beams of $1'$ and $3.4'$ at the respective frequencies.  A
full synthesis can achieve a continuum sensitivity (rms) of
0.28\,mJy/beam at 1420\,MHz and 3.8\,mJy /beam at 408\,MHz, and the
256-channel \ion{H}{i} spectrometer has an rms sensitivity of
3.5$B^{-0.5}\sin\delta$~K per channel, for total spectrometer
bandwidth $B$\,MHz and declination $\delta$.  The tuning range of the
telescope permits studies of Galactic and nearby extragalactic
objects.  The array uses 9\,m antennas, which provide very wide fields
of view of 3.1\degr\ and 9.6\degr\ (at the 10\% level), at the two
frequencies, and also allow data to be gathered on short baselines,
yielding extremely good sensitivity to extended structure.
Single-antenna data are also routinely incorporated into images to
ensure complete coverage of emission on all angular scales down to the
resolution limit.  In this paper we describe the telescope and its
receiver and correlator systems in detail, together with calibration
and observing strategies that make this instrument an efficient survey
machine.

     \keywords{Radio telescopes --
                aperture synthesis --
                wide-field imaging --
                \ion{H}{i} spectroscopy}
                
   \end{abstract}

%__________________________________________________________________________

\section{Introduction}

The Synthesis Telescope at the Dominion Radio Astrophysical Observatory
(DRAO) is a radio telescope with unique capabilities for Galactic
interstellar medium (ISM) studies.  Operating simultaneously at the
frequency of the spin-flip spectral line of atomic hydrogen (the \ion{H}{i}
line near 1420\,MHz), and in two continuum bands near 1420\,MHz and
408\,MHz, the telescope achieves arcminute angular resolution with
exceptional sensitivity to extended structure over wide fields (3.12\degr\
at 1420\,MHz and 9.63\degr\ at 408\,MHz at the 10\% level). These features
allow it to map several major constituents of the ISM, namely the atomic
gas (through the \ion{H}{i} line), the ionized gas (thermal continuum
emission detected in the continuum bands), and the relativistic component
(which generates synchrotron emission, measured in the continuum bands) in
a way which few other telescopes can match.  Its principal project at the
time of writing is the Canadian Galactic Plane Survey (\cite{Taylor00});
data from this survey are now entering the public domain.  The purpose of
this paper is to acquaint the scientific community with the characteristics
of the telescope.

The telescope in its original form was described by Roger et~al.\
(\cite{Roger73}), with the first ast\-ron\-omical res\-ults being
pub\-lished shortly there\-after (\cite{Costain76}; Ro\-ger \& Costain
1976).  Over the following two decades the capabilities of the
telescope were enhanced through a series of modifications, including
increasing the number of antennas from 2 to 4 and doubling the maximum
baseline in 1982, adding a 408\,MHz continuum channel in 1984, and
adding a second receiver path to the 1420\,MHz continuum system in
1986 to allow both hands of circular polarization (CP) to be measured.

An extensive rebuilding of the telescope began in 1992, to create a
telescope optimized for studies of the Galactic ISM, concentrating on high
angular resolution, wide-field studies of \ion{H}{i}.  Enhancements over
the following 3 years included the addition of three antennas, a new
1420\,MHz continuum correlator with double the reception bandwidth and
polarimetry capabilities, and a new \ion{H}{i} spectrometer with double the
number of channels.  The result is a substantially new telescope, built on
the basic infrastructure of the old.  In this paper we describe the
telescope, with emphasis on its new components.

\section{Array design}

\subsection{Basic design objectives}

The telescope was conceived as a tool for the study of the Galactic
\ion{H}{i}, which is known to have structure on all angular scales at which
observations have been attempted, from tens of degrees (\cite{Hartmann97})
to tens of milli-arcseconds (\cite{Faison99}).  An angular resolution of 1
arcmin was the target, an order of magnitude improvement over the
resolution available with the largest single-antenna radiotelescopes, while
retaining sensitivity to intermediate and larger structures. This required
a finely sampled aperture plane and short interferometer baselines, which
in turn dictated the use of small antennas. It was also clear at the outset
that it would be necessary to incorporate single-antenna data into the
Synthesis Telescope images (see \S2.3) and to mosaic individual images
together to extend the field of view.

A telescope which can successfully image \ion{H}{i} emission will also be
suitable for imaging Galactic continuum emission, which similarly displays
a wide range of structural sizes. Continuum emission at decimetre
wavelengths originates in the ionized and relativistic components of the
ISM. The latter gives rise to a small percentage of linear polarization in
the emission, and the ability to measure polarization was included in the
telescope to give information on magnetic fields at the point of origin and
on the magneto-ionic medium along the line of sight.

With only seven antennas, the number of instantaneous baselines is just 21;
many more baselines are required to meet the desired imaging criteria.  The
chosen configuration was therefore an east-west aperture synthesis
telescope with movable elements. Earth rotation varies the baseline vector
during an observation, and the movable antennas are relocated until the
required set of baselines has been completely sampled in consecutive
observations. (The normal observing strategy for the telescope is described
in \S8).

\subsection{Array configuration}

The task of designing an array configuration to meet the above objectives
was constrained by the existing four-antenna array, which had two movable
antennas on 300\,m of rail, along with two fixed antennas, one at the west
end of the rail and one $\sim600$\,m to the east.  With the possibility of
adding only three additional antennas, the final configuration (Figure~1)
was dictated by the needs of both continuum and \ion{H}{i}-line imaging
described above.

\begin{figure}
\resizebox{8.5cm}{!}{\includegraphics{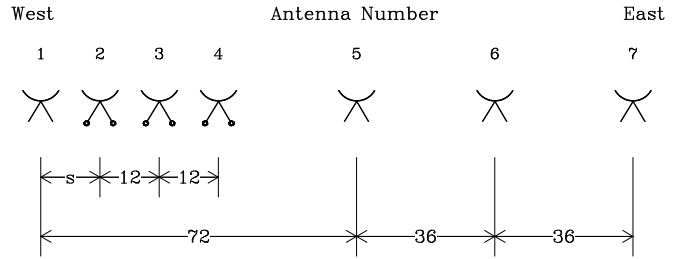}}
\caption{ The telescope configuration. Antenna separations are shown
in units of $L=4.286$\,m. The separation, s, between antennas 1 and 2
varies from $3L$ to $14L$ as antennas 2, 3, and 4 are moved along a
rail track.}
\label{fig-1}
\end{figure}

The diameter of the antennas used in the array is 8.5\,m (for further
details see \S 3). The baseline increment is $L=4.286$\,m, about half
an antenna diameter. This provides the necessary fine baseline sampling,
and places the first grating response at an angular radius where the
primary beam of the antennas is at a very low level, so that grating
responses to objects within the usable field of view lie outside the field
(although responses to strong sources outside the usable field can still
appear within it).  Antenna shadowing and the risk of mechanical
interference limit the minimum usable baseline to $3L \simeq 12.86$\,m,
which is sufficient to allow full representation in the images of
structures as large as 40~arcmin at 1420\,MHz and 2.3\degr\ at 408\,MHz.  A
maximum baseline of $144L \simeq 617.1$\,m provides the desired 1-arcmin
resolution limit at 1420\,MHz (at 408\,MHz a resolution of 3.4~arcmin is
achieved).  Table~1 lists the telescope specifications.

\begin{table*}
\caption[]{Telescope specifications}
\label{tabl-1}
\begin{center}
\begin{tabular}{lll}
%\hline
Operating frequencies:      &              & 1420\,MHz \\
                            &              & 408\,MHz \\
Number of antennas:         &              & 7 \\
Antenna diameter:           &              & five \@ 8.53\,m, two \@ 9.14\,m \\
Maximum baseline:           &              & 617.18\,m \\
Minimum baseline:           &              & 12.86\,m \\
Baseline increment:         &              & 4.286\,m \\
Visibility averaging time:  &              & 90 seconds \\
Field of view:              & 1420\,MHz    & 2.65\degr~diameter to 20\% \\
                            & 408\,MHz     & 8.22\degr~diameter to 20\% \\
Angular resolution:         & 1420\,MHz    & $58'' \times 58''$\,cosec$\delta$ \\
                            & 408\,MHz     & $3.4' \times 3.4'$\,cosec$\delta$ \\
Radius of first grating ring:  & 1420\,MHz & 2.82\degr \\
                               & 408\,MHz  & 9.82\degr \\
Polarization imaging:       & 1420\,MHz    & Stokes $I$, $Q$, and $U$ \\
                            & 408\,MHz     & One hand of circular \\
                            &              & (usually RHCP) \\
System temperature:         & 1420\,MHz    & 60\,K \\
                            & 408\,MHz     & 105\,K + $T_{sky}$ \\
Continuum bandwidth:        & 1420\,MHz    & 30\,MHz  \\
                            & 408\,MHz     & 3.5\,MHz \\
%Continuum sensitivity:      & 1420\,MHz    & 0.23 mJy/beam rms \\
%                            & 408\,MHz     & 3.3 mJy/beam rms \\
Tuning range for \ion{H}{i}:  &            & $-$1100 to +3000 km~s$^{-1}$ \\
Spectrometer frequency coverage ($B$): &   & 0.125, 0.25, 0.5, 1.0, 2.0, 4.0\,MHz \\
Number of spectrometer channels &          & 256 \\
Velocity coverage for \ion{H}{i}: &        & 211$B$ km~s$^{-1}$ \\
Channel separation:         &              & 0.824$B$ km~s$^{-1}$ \\
Channel width:              &              & 1.32$B$ km~s$^{-1}$ \\
Noise on 1-channel spectral map: &         & $3.5 B^{-0.5}\sin\delta$\,K \\
%\hline
\end{tabular}
\end{center}
\end{table*}

The three movable antennas are mounted on motorized platforms that ride on
a precise, machined track that is dimensionally stable and accurately
surveyed. Station markers along the track are used in positioning the
antennas, with a one-station move taking about 5 minutes.  The antennas are
usually moved so that they are $12L$ apart, with the separation between
antennas 1 and 2 ($s$ in Fig. 1) ranging from $3L$ to $14L$.  By this
means, complete coverage of baselines from $3L$ to $141L$ (plus $144L$) is
obtained with 12 settings of the movable antennas.  The time required to
synthesize a complete aperture is therefore $12\times12$ hours.  The
telescope is almost always used in this mode.  An exception is solar
imaging (Burke \& Tapping \cite{Burke95}); for this purpose a set of
spacings has been selected which allows the formation of a satisfactory
image of the Sun (in the absence of rapid variation such as burst activity)
in one 12-hour observation.

For \ion{H}{i} imaging, only the 12 baselines between the four fixed and
three movable antennas are used.  \ion{H}{i}-line images are typically of
limited dynamic range, determined by the ratio of the maximum brightness
temperature (rarely more than 120\,K) to the expected noise level
($\sim3$\,K).  However, antenna-based calibration parameters, determined
from continuum measurements using all 21 baselines, are routinely applied
to \ion{H}{i} visibilities.

\subsection{Complementary single-antenna data}

The short baselines available on the Synthesis Telescope permit accurate
imaging of quite large features, but it is still necessary to incorporate
data from single-antenna radio telescopes into the images. The DRAO 26-m
Telescope is usually used to provide the necessary information on the
largest \ion{H}{i} structures.  Single-antenna continuum data at 1420\,MHz
are obtained from the Effelsberg surveys of Kallas \& Reich
(\cite{Kallas80}), Reich et~al.\ (\cite{Reich90}), and Reich et~al.\
(\cite{Reich97}) or from the Stockert surveys of Reich (\cite{Reich82}) and
Reich \& Reich (\cite{Reich86}).  When a long-wavelength channel was
planned for the Synthesis Telescope in the 1980s, the frequency of 408\,MHz
was chosen because of the existence of a complementary, well-calibrated,
single-antenna survey of the whole sky at that frequency (\cite{Haslam82}).

\section{Antennas}

The antennas used in the Synthesis Telescope have equatorially mounted,
paraboloidal reflectors.  There are two slightly different designs in use,
the most significant difference being that two antennas (at the east and
west extremes of the array) have diameter $d=9.14$\,m and focal length
$f=3.81$\,m ($f/d=0.42$), while the remainder have $d=8.53$\,m and
$f=3.66$\,m ($f/d=0.43$).  The difference in antenna diameters leads to a
mismatched weighting of {\it u-v} samples which is sufficiently large to
affect imaging; Willis (\cite{Willis99}) has devised a means of
compensating for this effect.

Signals are collected by a dual-frequency feed (Veidt et~al.\
\cite{Veidt85}; Trikha et~al.\ \cite{Trikha91}) mounted at the prime focus.
Circular polarization (CP) is received at both frequencies: continuum
emission is expected to have a small fraction of linear polarization (from
synchrotron radiation), and measurement of that component can be made most
simply by using CP feeds.

At 1420\,MHz a multi-mode feed, based on one of the designs of Sheffer
(\cite{Scheffer75}), provides a flat-topped illumination pattern, yielding
an aperture efficiency $\eta_A=0.55$, with an edge illumination of the
reflector of $-16$\,dB.  A ``quarter-wave plate'' in the waveguide
(consisting of three reactive posts; \cite{Simmons52}) provides a 90\degr\
phase shift, and linear probes aligned at 45\degr\ to the plane of the
quarter-wave plate collect left-hand circular polarization (LHCP) and
right-hand circular polarization (RHCP) simultaneously.

The circularly polarized 408-MHz feed was added subsequently, and was
designed to avoid any degradation of 1420-MHz performance.  The outputs of
four orthogonal monopoles, placed in the outer cavity of the 1420-MHz
feed, are combined in a TEM-mode hybrid.  A switch selects either LHCP or
RHCP; normally RHCP is received (the use of a CP feed gives an accurate
measurement of Stokes $I$ even in the presence of some linearly polarized
emission).  At 408\,MHz the feed has the radiation characteristics of a
circular waveguide of diameter $0.7\lambda$, illuminating the aperture very
broadly.  The aperture efficiency is $\eta_A=0.6$, but edge illumination is
high ($-8$\,dB) and beam efficiency is low ($\eta_B=0.70$) because of the
high spillover.

Antenna noise temperature at 1420\,MHz varies from a high of 26.0\,K on one
antenna to a low of 17.8\,K on another, because of slight differences in
construction.  Table~2 gives details of the noise budgets of the best and
worst antennas.  These data were derived by Anderson et~al.\
(\cite{Anderson91}) based on two-dimensional mapping of the radiation
pattern of one of the antennas.  The feed-support struts of several
antennas have been modified (Landecker et~al.\ \cite{Landecker91}) to
reduce the ground noise which the struts scatter into the aperture.

\begin{table}
\caption[]{Antenna noise-temperature budget at 1420\,MHz (K)}
\label{tabl-2}
\begin{center}
\begin{tabular}{lllll}
%\hline
Cosmic microwave background    & 2.7   &     &      & \\
Galactic emission              & 1.0   &     &      & \\
Atmospheric emission (zenith)  & 2.0   &     &      & \\
{\bf Main beam contribution:}  & {\bf 5.7} &     &      & \\
                               &       &     &      & \\
                               & {\it Worst} &     & {\it Best} & \\
                               & {\it antenna} &   & {\it antenna} & \\
Spillover                      & 8.0   &     & 8.0  & \\
Diffraction at reflector rim   & 0.6   &     & 0.6  & \\
Mesh leakage                   & 5.9   &     & 1.5  & \\
Strut scattering               & 5.8   &     & 2.0  & \\
{\bf Sidelobe contributions:}  & {\bf 20.3} &     & {\bf 12.1} & \\
{\bf Total:}                   & {\bf 26.0} &     & {\bf 17.8} & \\
%\hline
\end{tabular}
\end{center}
\end{table}

\section{Receivers}

\subsection{The 1420-MHz receiver}

Sensitivity at 1420\,MHz is primarily determined by the low-noise HEMT
amplifier, which has a noise temperature, $T_R \approx 35$\,K (for 16
amplifiers built, 28\,K $< T_R < 38$\,K and $\langle T_R \rangle=33.4$\,K).
The uncooled 3-stage amplifier is an enhanced version of the amplifier
described by Walker et~al.\ (\cite{Walker88}).  It uses source-inductance
feedback to achieve a good input match (Weinreb et~al.\ \cite{Weinreb82}).

After filtering to suppress the image band, the signal is converted to an
intermediate frequency (IF) band of 12.5--47.5\,MHz.  IF signals are
conveyed from each antenna to the central building by coaxial cables.  No
compensation is attempted for changes in IF phase due to cable-length
changes with temperature; physical matching of cables provides adequate
stability.

The Local Oscillator (LO) system (Landecker \& Van\-el\-dik \cite{Landecker82})
delivers signals of controlled frequency and phase to the mixers at each
antenna. To control phase, the system uses the ``round-trip'' design, in
which the electrical length of the cable connecting the central electronics
to the antenna is measured by sending a signal out to the antenna and back.
If the cable electrical length changes by $\epsilon$, the returning signal
changes its phase by the equivalent of $2\epsilon$.  Closing the feedback
loop requires a division by 2; consequently there is an ambiguity of $\pi$
in the output phase.  This difficulty is overcome by operating the
round-trip system at half the frequency of the required LO signal.  At the
antenna, the output is doubled to produce the LO signal near 1390\,MHz.
The tuning range of the LO system is given in Table~1; it accommodates
observations of \ion{H}{i} in the Galaxy and in nearby extragalactic
systems.

The interferometer fringe rate is reduced to zero for each baseline by
rotating the phase of the LO signal for each antenna relative to the
reference antenna (Antenna 5 at the physical centre of the array -- see
Figure~1).  Phase switching in a Walsh-function pattern (Granlund et~al.\
\cite{Granlund78}) is also applied to the LO signal to reduce spurious
correlation from crosstalk between receiver channels. The switching pattern
has 16 steps of 5.6 s in the telescope visibility-averaging time of 90 s.

Compensating delays are inserted in the IF signal paths to equalize path
lengths, from the incoming wavefront, through each antenna to the
correlator.  Stripline paths are used for short delays and coaxial cable
for longer delays, selected by PIN diode switches.  Cable delay sections
are equalized for attenuation and dispersion.  This is an enhancement of
the earlier system described by Landecker (\cite{Landecker84}).  The
minimum delay step of 3.91\,cm ($\approx 0.13$\,ns) is determined by
the specification that the delay error should produce a phase difference
across the spectrometer band of less than 0.1\degr. Delay system
specifications are given in Table 3.

\begin{table}
\caption[]{1420\,MHz delay system specifications}
\label{tabl-3}
\begin{center}
\begin{tabular}{lll}
%\hline
Delay medium: & stripline and coaxial cable \\
Frequency range: & 10 to 50\,MHz \\
Delay step:  & $2^{-8}\lambda_{30} = 3.91$\,cm \\
Total delay: & $64\lambda_{30} = 640$\,m \\ 
Amplitude accuracy: & $\pm0.02$\,dB (rms) \\
Phase accuracy: & $\pm1$\degr (rms) \\
%\hline
\end{tabular}
\end{center}
\end{table}

Residual gain fluctuations due to imperfect delay-cable equalization or
unequal switch losses are removed by an automatic gain control (AGC).  The
AGC measures signal levels at the delay-system input and output, and keeps
the net gain of the delay system constant.  A second gain monitor spans the
entire signal path, from antenna feed to correlator output: small modulated
noise signals ($\Delta T_A \approx 1$\,K) are injected into each antenna
feed, and are demodulated from autocorrelation outputs (see \S 5.1) to give
continuous measurements of system gain and system temperature.

Since this is a spectroscopic telescope, the reception frequency must be
varied to account for diurnal and orbital motion of the Earth, as well as
for the radial velocity of the source under observation.  The total
reception band, of width 35\,MHz, is kept centred on the \ion{H}{i} band
selected for observation by continuously tuning the first LO.  The
continuum reception band is divided into four sub-bands distributed on
either side of the spectroscopy band, as illustrated in Figure~2.  In the
course of an observation, continuum band centres vary as the LO is tuned,
but the imaging software accounts for this slight change. The five
reception bands are translated to fixed locations in the IF band.
 
\begin{figure}
\resizebox{8.5cm}{!}{\includegraphics{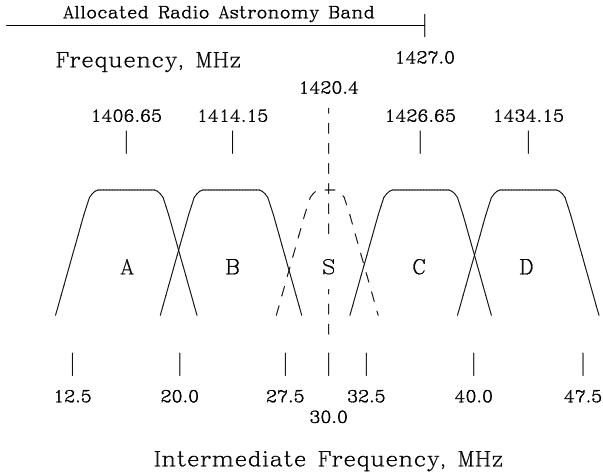}}
\caption{The arrangement of continuum and spectroscopy bands near
1420\,MHz.  S is the band used for \ion{H}{i} spectroscopy, and A, B, C,
and D are continuum bands. The system LO is varied to keep the \ion{H}{i}
information centred in the IF at 30.0\,MHz. The band allocated to radio
astronomy extends from 1400 to 1427\,MHz; part of band C and all of band D
are outside the protected band. The overall width of the spectroscopy band
can be varied from 0.125 to 4\,MHz (see text).}
\label{fig-2}
\end{figure}

The four continuum bands are defined by IF bandpass filters and are
converted to baseband using fixed second LOs. A quadrature replica of each
band is created using a phase-shifted second LO.  The spectrometer band is
converted to lower frequency in a single-sideband mixer. The overall
bandwidth of the spectrometer is selectable with six options from
$B=0.125$\,MHz to $B=4.0$\,MHz in multiples of 2. Before digitizing, the
band is sharply defined by a six-pole elliptic-function bandpass filter
spanning $B$ to $2B$. Use of this ``quasi-baseband'' allows better image
rejection in the single-sideband mixer.

\subsection{The 408-MHz receiver}

The 408\,MHz receiver is designed to use as much as possible of the band
assigned to radio astronomy (406.1--410\,MHz), while suppressing
out-of-band communications signals.  Several stages of filtering are
distributed between gain stages in the receiver, and a total bandwidth of
3.5 MHz is achieved.

A single coaxial cable carries both LO and IF signals between the focus of
each antenna and the central receiver building.  The LO phase is controlled
by a feedback system which keeps the phase of the 30\,MHz IF signal
arriving at the centre of the telescope completely independent of the
length of the interconnecting cable (Veidt et~al.\ \cite{Veidt85}). All
subsequent signal processing is carried out in digital electronics or in
software (see \S 5.3).  A minor drawback of the LO-IF system is the
generation of an unwanted fixed-frequency signal at the centre of the IF
band; this is removed prior to correlation by a narrow bandstop filter.

\section{Correlators}

The three operating bands of the telescope, the 1420-MHz continuum band,
the \ion{H}{i} spectroscopy band, and the 408-MHz continuum band, use
digital correlators of different designs, which we label the C21, S21, and
C74 systems respectively.  The three correlators were built at different
times; their design differences are partly dictated by that history but are
mostly driven by the significantly different requirements. Table~4 lists
the specifications of the three correlators.

\begin{table*}
\caption[]{Correlator characteristics}
\label{tabl-4}
\begin{center}
\begin{tabular}{llll}
%\hline
                  &                  &            &            \\
{\bf Correlator}  & {\bf C21}        & {\bf S21}  & {\bf C74}  \\
                  &                  &            &            \\
Bandwidth (MHz)   & 7.5 ($\times 4$) & $2^n$, $-3 \leq n \leq 2$ & 4.0$^*$ \\
Sampling rate (MHz) & 20.0           & 2$f_N^\dagger$ & 16.0 \\
Correlator efficiency & 0.985        & 0.88       & 0.88       \\
Number of quantizer levels & 14 & 3 & 3 \\
                  &                  &            &            \\
\multicolumn{4}{l}{$^*$ Bandwidth actually used is 3.5\,MHz, defined by 
RF filters} \\
\multicolumn{4}{l}{$^\dagger f_N=$ Nyquist~frequency}    \\
%\hline
\end{tabular}
\end{center}
\end{table*}

Digital representation of signals introduces quantization noise, some of
which can be recovered by increasing the sampling rate (Bowers \& Klingler
\cite{Bowers74}; Klingler \cite{Klingler74}). Figure~3 shows correlator
efficiency for the two quantization schemes used in this telescope.

\begin{figure}
\resizebox{8.5cm}{!}{\includegraphics{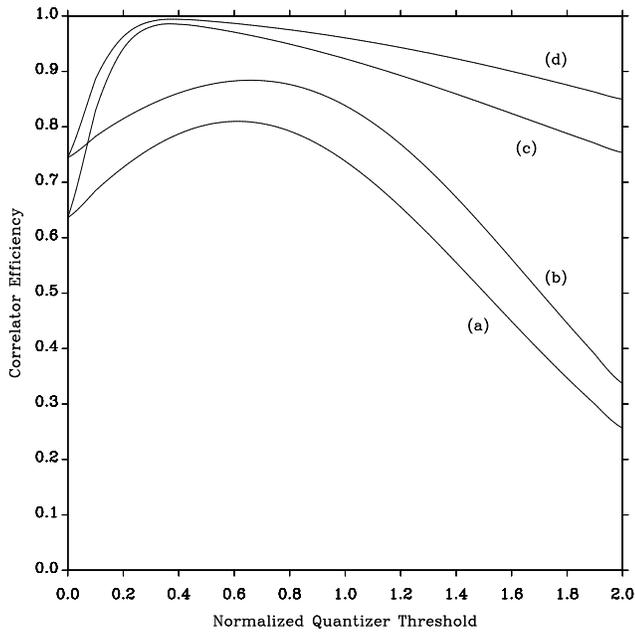}}
\caption{Correlator efficiency. Curves (a) and (b) show the efficiency of
a 3-level correlator at Nyquist and double-Nyquist sampling rates
respectively, and (c) and (d) show the same for a 14-level correlator. The
horizontal axis is labelled with the voltage setting of the first quantizer
threshold in units of the r.m.s. value of the input signal. In each
correlator on the telescope, the input signal level is chosen to optimize
correlator efficiency.}
\label{fig-3}
\end{figure}

Sections 5.1, 5.2 and 5.3 describe the three correlators in turn.  Section
5.4 describes routines developed to correct the inherent correlator
non-linearity.  These corrections are currently applied to C21 and S21 data
only.

\subsection{The 1420-MHz continuum correlator}

The C21 correlator (\cite{Karpa89}) uses fourteen-level representation of
the input signals; with moderate over-sam\-pl\-ing, it achieves an unusually
high correlator efficiency ($\eta_c=0.985$; see Figure~3).  The correlator
was designed to provide products from all antenna pairs, and to form all
four polarization products from the LHCP and RHCP inputs. However, the
total continuum bandwidth is too wide to permit mapping of the entire field
of view of the telescope without bandwidth smearing (decorrelation arising
from differential delay of off-axis, wide-bandwidth signals; see
\cite{Bridle89}). The 30\,MHz band is therefore divided into four 7.5\,MHz
sub-bands as illustrated in Figure~2.  Each sub-band input is digitized to
14 levels and the products are formed in one of four identical correlators.

Input signals are digitized using four-bit monolithic flash converters
(Analog Devices AD9688) with the sampling clock running at 20\,MHz,
slightly faster than the Nyquist rate for a 7.5-MHz bandwidth.  Of the 16
possible A/D output levels, only 14 are used.  The two extreme levels
(outputs 0000 and 1111) are reserved for use as control signals embedded in
the data stream.

Each correlator module uses a ROM look-up table where an 8-bit product is
stored at the address specified by the two 4-bit inputs.  The product is
accumulated for 5.6 seconds.  The occurrence of the code 0000 at either
input of the multiplier causes accumulation to stop.  The accumulator (a
combination of synchronous and ripple counters) is allowed to settle, and
is then read by the system microprocessor.

The occurrence of code 1111 on one of the multiplier inputs causes the
multiplier to produce an 8-bit output corresponding to the square of the
4-bit number appearing on the other input.  Extra correlator cards are used
in this way as auto-correlators, and they generate outputs corresponding to
the power in each input signal.  Small noise signals are injected into each
antenna (see \S4.1), and the autocorrelation outputs are used to measure
both system gain and noise in each channel. Additional correlator channels
are used to compute the mean of each input signal (by multiplying the input
signal by a fixed number), a measure of the error in each A/D converter.
Similarly, ``self-correlations'' are formed between in-phase and quadrature
signals from a given antenna to measure any deviation from orthogonality
(which is corrected for in subsequent processing).

The flow of input signals from the A/D converters to the correlators, and
the flow of output data from the correlators, is handled by the
Control/Interface (CI) modules.  These modules include a $4096 \times
4$-bit RAM, located at the boundary between data acquisition and data
processing, which serves two purposes. The RAM can store A/D converter
outputs for later analysis (used in testing A/D converters).
Alternatively, it can be loaded from the microprocessor with data which can
be passed to the correlators in place of the normal input signals.  In this
way all correlator channels can be subjected to a test which uses realistic
input signals while giving exactly known outputs.  The entire correlator
system undergoes self-test in this fashion at the start of every
observation.

The correlator modules are arranged in a matrix.  Signals move one step
along the rows and diagonals of the matrix at each clock cycle and are
re-timed at each correlation site.  Since control signals are embedded in
the data stream, timing is very simple, and the correlation matrix is (in
principle) extendible to handle any number of antennas.

\subsection{The \ion{H}{i} correlation spectrometer}

The S21 correlator forms only those baseline products needed for complete
sampling of the {\it u-v} plane, the 12 combinations of fixed and movable
antennas (Figure~1).  For maximum sensitivity, two identical correlators
form products from LHCP and RHCP signals.  Cross-correlation between
signals from the two hands of polarization is not currently available.
This correlator design, in auto-correlator mode, is also used on the DRAO
26-m Telescope for observation of large-scale \ion{H}{i} structure for
combination with Synthesis Telescope images.

The S21 correlator uses many modules from the C21 system, including the A/D
converter, the CI modules, the backplanes, and the microprocessors.  Many
functions from the C21 system are reproduced, including the autocorrelation
circuits and the circuits used to compute means of input signals (see \S
5.1).  The significant difference between the systems is the correlation
module itself.

Data streams, which leave the A/D converters encoded to 14 levels, are used
with 14-level precision for auto-correlation and other functions, but are
converted to 3-level coding for processing by the correlation modules.  The
14-level auto-correlators provide accurate measurement of input power
levels, and this information is used to linearize the cross-correlation
functions formed from 3-level data (see \S5.4).

The correlator was designed around a CMOS appl\-ic\-at\-ion-specific integrated
circuit (ASIC) developed for this project (Hovey \cite{Hovey98}). The ASIC
computes a four-point correlation function from input signals quantized to
3 levels, and 128 ASICs implement a 512-lag correlator (256 complex
channels) for each baseline.  The ASIC architecture is flexible, allowing
both cross- and auto-correlation of data sampled at the Nyquist rate,
$f_N$, or at $2f_N$.  Sampling at $2f_N$ is used because it increases
correlator efficiency from 0.81 to 0.88.

\begin{figure}
\resizebox{8.5cm}{!}{\includegraphics{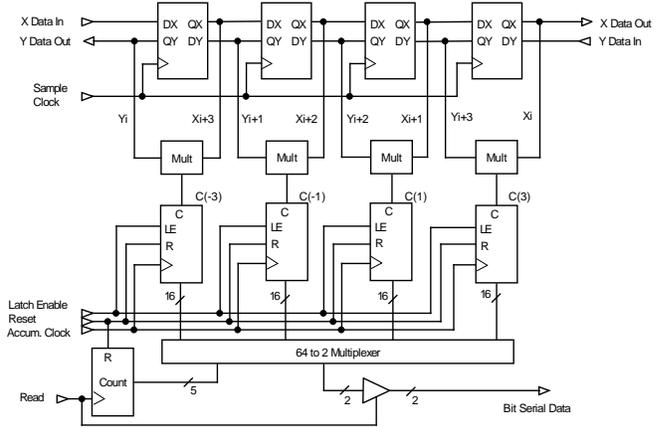}}
\caption{Functional diagram of the ASIC developed for the spectral-line
correlator (the S21 system).}
\label{fig-4}
\end{figure}

Figure~4 shows the organization of the ASIC.  The $x$ and $y$ inputs are
3-level quantities, taking values $(1, 0, -1)$.  The shift registers are
arranged so that $x$ and $y$ streams flow in opposite directions.
Correlation function values are formed by connecting a
multiplier-accumulator across each shift-register output.  The multiplier
output is biased, so that only positive products $(0, 1, 2)$, are formed,
allowing accumulation in a simple up-counter.  Accumulation begins with a
2-bit adder, whose output synchronously clocks a 23-bit ripple counter.
Adder outputs are clocked into this counter at a fixed rate so that the
mean value of the accumulator is constant even though the sample rate is
adjusted when the overall bandwidth is changed.  The system microprocessor
detects overflows of the most significant accumulator bit, and extends the
accumulation interval to 5.6 s.

The cross-correlation function is under-sampled in the ASIC because the
delay interval is $2\tau$, where $\tau$ is the period of the sampling
clock.  By adding a second ASIC with one input delayed by $\tau$, the
correlation function can be fully sampled.  Alternatively, if the input
signals are sampled at $2f_N$, a fully sampled correlation function is
obtained (this is the normal mode of operation).  For auto-correlation, a
correlation function with one point at zero lag can be created by delaying
one input by one half-lag.

\subsection{408-MHz continuum correlator}

The C74 correlator was designed to carry out interferometer
signal-processing operations in simple software procedures to the maximum
extent feasible (Lo et~al.\ \cite{Lo84}).  Figure~5 is a diagram of the
system, showing the distribution of signal-processing functions between
hardware and software.

\begin{figure}
\resizebox{8.5cm}{!}{\includegraphics{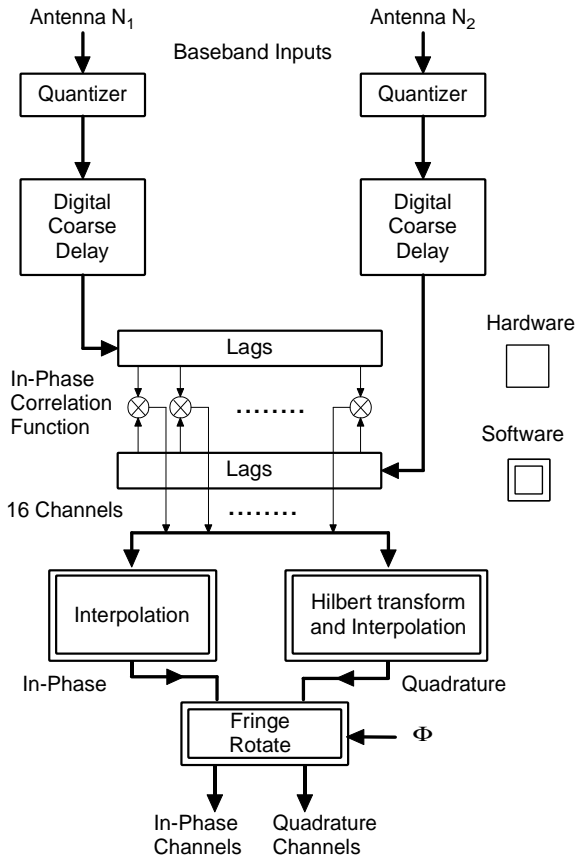}}
\caption{Functional diagram of the correlator used with the 408\,MHz
continuum channel of the telescope (the C74 system). The division of
functions between hardware and software is indicated.}
\label{fig-5}
\end{figure}

Incoming signals are sampled at 16\,MHz and coarse delay is inserted in
steps of one cycle of this clock (62.5\,$\mu$s).  A 16-point correlation
function is generated.  The in-phase visibility is retrieved from this
function by interpolating to the exact geometrical delay.  The quadrature
visibility, which contains independent information, is also derived from
this function. It is calculated by convolving the in-phase correlation
function with the kernel function of a band-limited Hilbert transform.
Since the in-phase and quadrature visibility outputs are required at only a
single value of delay, the convolution degenerates to a simple dot product,
requiring multiplication of each value of the correlation function by a
matching coefficient from the kernel function.  This is a simple and quick
operation in the microprocessor (see Lo et~al.\ \cite{Lo84} for details).
A small error (of the order of 5\%) arises in the derivation of the
quadrature visibility because of truncation of the kernel.  The error is
mostly corrected by a multiplication factor.  The residual error is much
smaller than the decorrelation arising from bandwidth smearing.

Delay equalization is a continuous process, since the required path
compensation is continuously changing as the Earth rotates.  Precision of
the compensation is high, limited only by the numerical precision of the
interpolation arithmetic and the rate at which interpolation is performed.
This confers an advantage over systems where delay is inserted in discrete
steps.  Such steps affect the phase of the correlator output, and must be
corrected for, either before or after correlation.

\subsection{Correlator linearity}

Quantization of the baseband signals introduces non-lin\-ear\-it\-ies in the
correlator output, which can lead to substantial artefacts in images (both
\ion{H}{i}-line and continuum) when bright Galactic objects lie in the
field. Because of its sensitivity to extended structure, this telescope is
more prone to this problem than other synthesis telescopes which resolve
out most of the broad emission.

There are two approaches to the derivation of accurate visibilities using a
digital correlator. In one approach, the signal level at the input to the
A/D converter is held constant at the optimum level (see Figure~3). Under
these circumstances, a 3-level correlator yields a value of correlation
coefficient, $\rho$, linear within 1\% up to values of
$\rho=0.38$. However, to recover accurate visibility amplitudes requires
accurate tracking of variations of system noise temperature. The
alternative, which we have adopted, is to stabilize receiver gain through
the system but to allow the signal level at the A/D converter input to vary
(caused, for example, by variations of atmospheric emission and received
ground radiation as a function of antenna elevation angle). This approach
requires the development of algorithms to linearize correlator output.  The
algorithms convert the output of the correlators (separate algorithms have
been developed for the 3-level and 14-level correlators) into an accurate
measure of cross-power in the face of changes of system temperature of as
much as 30\%.  The algorithms must be applicable to both auto- and
cross-correlators, and must therefore be able to handle correlation
coefficients as high as $\rho=1.0$. However, $\rho$ can also be high in the
cross-correlator: when observing \ion {H}{i} emission with the 12.86-m
baseline, $\rho$ can rise to values as high as 0.7.

The algorithms proceed by first linearizing the autocorrelator response to
derive an accurate measure of the variance of the input signal. Using this
measurement, the cross-correlator response is linearized to obtain an
accurate value of cross-power.  The equations describing the correlator
output in terms of its inputs cannot be directly inverted. Instead we use
series approximations, following the ideas of Kulkarni \& Heiles
(\cite{Kulkarni80}) and D'Addario et~al.\ (\cite{D'Addario84}). However,
the correlator becomes increasingly non-linear as the input signal
increases, spending a greater fraction of time above the highest quantizer
decision level.  The coefficients of the series inversion formula therefore
must vary with input signal level.  In our algorithm, these coefficients
are themselves estimated by evaluating a series. Details of the
linearization algorithms can be found in Hovey (\cite{Hovey98}).  Their
application reduces errors of as much as 30\%, which occur in common
observing conditions with this telescope, to less than 0.1\%.

\section{Software}

The observing software comprises a combination of Unix-based tasks
running on an IBM RS/6000, and a set of software tasks which run on
Motorola-68000-based microprocessors (Figure~6).  The RS/6000 system
provides high-level synchronization, scheduling and data storage,
while the true real-time work of data collection and sampling is done
by the microprocessors.  After data are collected, they are
transferred via an ethernet connection to another IBM RS/6000 system,
where the data are processed through automatic error-detection and
flagging routines and inserted into a database awaiting final
calibration and imaging, which is carried out once all the data
required for a complete synthesis have been collected.  (Twelve
observations of 12 hours each are required to obtain all the data
needed for imaging; see \S8).

\begin{figure*}
\resizebox{12cm}{!}{\includegraphics{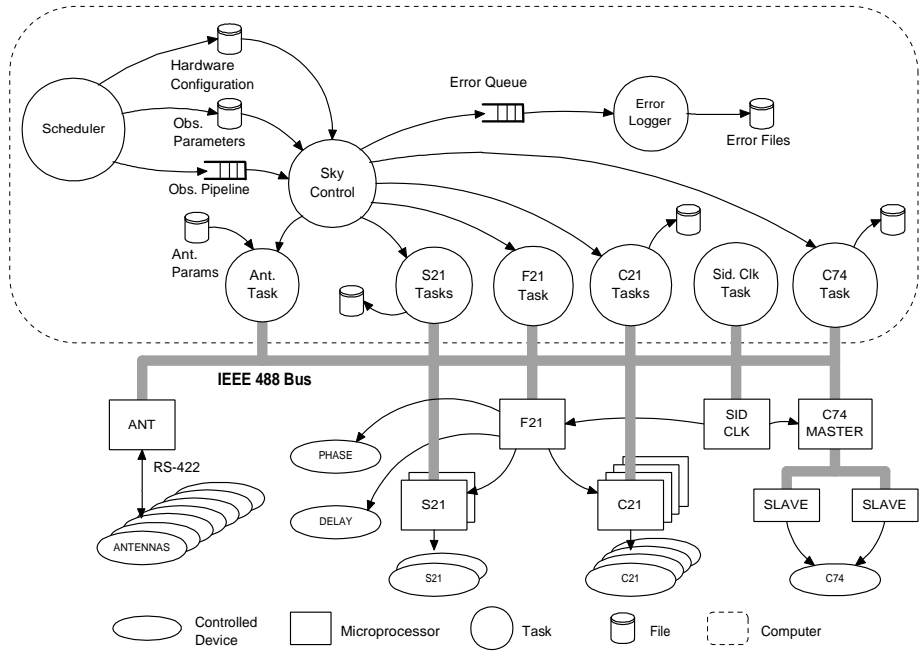}}
\hfill
\parbox[b]{53mm}{
\caption{The telescope control software. C21, S21 and C74 are the three
correlators described in section 5. F21 denotes the task and microprocessor
which control the 1420\,MHz receiver (frequency, phase, and delay). SID CLK
is the sidereal clock. At the completion of an observation all files
indicated are transferred via an ethernet connection to another computer
where further processing occurs.}
\label{fig-6}}
\end{figure*}

\subsection{Telescope Control Software}

The user interface to the observing system (Figure 6) is the program
\verb+SCHEDULER+, which allows the user to initiate observations of the
program field and calibration observations, check on their progress, and
abort them.  A ``template'' consisting of a list of observations specified
by field centre, duration, and starting time or hour angle may be entered,
with individual sub-systems selectively chosen for each.  \verb+SCHEDULER+
checks that the observations are possible within the antenna horizon
limits, and the list is then placed in the observing queue.

Observing-system tasks communicate via serial polling over an IEEE~488
interface with microprocessors embedded in the various sub-systems of
the telescope.  These tasks read the sidereal clock, log messages
generated by the observing system, and provide two-way communication
with the microprocessors.  Global synchronization between the
sub-systems is performed by the program \verb+SKY_+ \verb+CONTROL+
using Unix IPC messaging via addresses assigned to each task.

\verb+SKY_CONTROL+ also processes the observing queue. It determines which
sub-systems are needed, initializes control files, and signals sub-system
tasks accordingly.  It also creates a binary file containing all of the
parameters and hardware settings needed for the observation.  This file
subsequently follows the data through post-observation processing software
as a record of the state of the system when the data were acquired.
Sub-system-specific ASCII parameter files are also generated, which are
read by the controlling tasks and used to send start-up commands to the
microprocessors.

Once an observation has begun, \verb+SKY_CONTROL+ goes into a wait mode,
looking for either the end of the run, or an ``ABORT'' signal from
\verb+SCHEDULER+.  At the end of an observation \verb+SKY_CONTROL+ checks
the observing queue, and initializes the next observation, continuing until
the queue is empty.

Observing-system timing is controlled by the sidereal clock. Its attached
microprocessor sends the time, once per second, to a shared memory area
which can be accessed by other tasks.  The time is also directly
distributed to the individual microprocessors.

\subsection{Error Logging}

All Unix tasks report errors to a central program, \verb+LOGGER+, via an
IPC message queue. \verb+LOGGER+ runs continuously and writes all messages
to a terminal and into an ASCII output file.  Upon completion of an
observation, the file is closed, tagged to the observation, and a new file
is started.  This tagged file follows the observation data into the
post-observation processing area, and is used as input to the flagging
programs for initial editing of bad data.

\subsection{Post-Observation Processing}

When an observation is complete, \verb+SKY_CONTROL+ moves the data across
an ethernet link to an off-line post-observation processing area on disk,
and spawns sub-shells to run command procedures.  These are typically Unix
shell scripts which call a sequence of FORTRAN-based programs that each
perform a specific operation on the data. Important information and
statistics may be generated by these programs, and are written to a
processing log file, which the user checks in order to verify the health of
the system and the validity of the data.

Since Synthesis Telescope observing projects are typically spread over
several days, a number of project-specific databases are maintained, into
which each newly pro\-cess\-ed observation or calibration is inserted.  Each
entry is indexed with the date of observation and the interferometer
configuration.  At any stage of data gathering these databases may be
examined. However, aside from some manual flagging of bad data in each
observation, the data are not generally further processed until all 12
observations for the project are complete.

Once all the data are available, an operator uses a graphics-based
visibility editor to flag (but not remove) any remaining bad data.
Appropriate calibration parameters for amplitude, phase,
cross-polarization, and spectrometer passband-flattening, derived from
calibration observations, are extracted from global databases and stored in
a project-specific table. The data and accompanying tables are then
processed to remove bad data and apply calibration.  The resulting
databases are suitable for gridding and Fourier transforming into images.

Images from the Synthesis Telescope are generally analyzed using an
extensive suite of locally written software, designed to provide optimal
results from Synthesis Telescope data, taking into account known
instrumental effects. A detailed description of this suite of software is
given in \cite{Higgs96} and Willis (\cite{Willis99}).

\section{Polarimetry}

As described above, the C21 correlator measures cross-correlations of all
combinations of polarizations; 1420-MHz continuum polarimetry became
available on the Synthesis Telescope in 1993 (see Smegal et~al.\
\cite{Smegal97}).

The antennas of the Synthesis Telescope are equatorially mounted, with
feeds that cannot be rotated relative to the reflector, so the beam pattern
always has a fixed orientation on the sky. While this has many advantages,
it does complicate separation of spurious instrumental polarization from
source polarization.  Corrections for instrumental polarization must be
determined by observing sources with little or no intrinsic polarization.
The method by which these corrections are determined is discussed in detail
by Smegal et~al.\ (\cite{Smegal97}); put briefly, in the absence of an
actual polarized signal, the cross-polarization correlations provide
information about the orthogonality of the nominal LHCP and RHCP signals on
each antenna, also known as ``leakage''.  The leakage terms measured for
the Synthesis Telescope using 3C\,147 and 3C\,295 amount to between 1 and
5\% on most antennas, although one antenna has about 10\% leakage.  If left
uncorrected, these errors produce spurious instrumental polarization at the
field centre amounting to several percent of the total flux.  After
correction, the residual instrumental polarization at the field centre is
reduced by about an order of magnitude to 0.25\% of the total intensity.

To measure polarization angle it is necessary to measure the phase
difference between RHCP and LHCP channels on one antenna, an arbitrarily
chosen reference.  This is achieved by observing 3C\,286, a strongly
polarized, compact source with a fractional polarization of 9.25\% and
polarization angle of 33\degr\ at 1420\,MHz.  Polarimeter accuracy is
5\degr\ in polarization angle, and 10\% in polarized intensity.  The
precision of the polarization angle measurements is also limited by
ionospheric Faraday rotation, which amounts to about 3\degr\ in
polarization angle between day and night (at present no correction is
applied).

The off-axis polarization properties of the instrument are very good, and
useful results can be obtained up to $90'$ from the field centre.  The
residual instrumental terms rise approximately quadratically from the field
centre, reaching about 10\% of total flux at this radius.  Empirical
corrections, determined by observing 3C\,147 and 3C\,295 at various
positions in the field and interpolating to intermediate positions, are
applied to remove this increase. These corrections have been measured three
times over a three-year period; they are very stable.

At present, deviations from circularity of the feeds are not accounted for.
Since the data are processed as if the polarizations were purely circular,
a small amount of any incident linear polarization is actually ascribed to
circular polarization (Stokes $V$).  This has little effect on linear
polarimetry, but severely limits the ability of the telescope to detect and
measure circular polarization. This has minimal impact on the astrophysical
questions addressed by the telescope. For details see Smegal et~al.\
(\cite{Smegal97}).

\subsection{Rotation Measure}

An important aspect of polarization studies is determining rotation measure
to de-rotate observed polarization angles and recover the orientation of
the intrinsic magnetic field.  The sensitivity of the Synthesis Telescope
to Faraday rotation is thus an important consideration.  At 1420\,MHz, a
rotation measure of 4\,rad\,m$^{-2}$ is required to produce a change in
observed polarization angle of 10\degr, which is the limit of detectability
with the present system.

In an image made from all 4 available bands, a rotation measure of
approximately 330\,rad\,m$^{-2}$ would cause 10\% depolarization, with
complete depolarization occurring for a rotation measure of
$1.2\times10^3$\,rad\,m$^{-2}$.  In a single band these figures are over 5
times higher, $1.7\times10^3$\,rad\,m$^{-2}$ and
$6.6\times10^3$\,rad\,m$^{-2}$, respectively.

Since the 4 bands are available for separate processing, it is possible to
use the frequency-dependent polarization angle to derive rotation measure.
However, the relatively small span of frequency available imposes fairly
high limits on the detectable rotation measures: 100\,rad\,m$^{-2}$ is
required to produce a change in angle of 10\degr\ between the centres of
bands A and D (see Fig. 2).

\section{Observing strategies}

The east-west nature of the Synthesis Telescope requires that an object be
tracked for 12 sidereal hours to obtain full hour-angle coverage. We
describe this as an {\it observation}, although the total duration may
sometimes be less than 12 hours.  Twelve such observations produce a
complete set of visibilities, fully sampling the available baselines; we
describe such a set of data as an {\it observation set}.

In normal use the telescope is scheduled in a 4-day cycle, during which 7
back-to-back 12-hour observations with interleaved calibrations are
completed, leaving approximately 8 hours per cycle for moving antennas and
maintenance.  The cycle is repeated 12 times to gather data on all possible
spacings for the 7 fields.  Occasional equipment failure and additional
maintenance can add an extra day per field; an observing rate of 46 to 52
fields per year can be achieved in this mode of operation.

\subsection{Calibration}

Amplitude calibration and phase referencing are achieved by observing a set
of strong, compact calibration sources which are known not to be time
variable.  The derivation of calibration coefficients uses standard
antenna-based algorithms, with the complication that, since the Synthesis
Telescope has a very wide field-of-view, it is necessary to account for
nearby extraneous sources, which would otherwise cause undesirable
hour-angle (and possibly time) dependencies.  To this end, visibility
models of the region surrounding each calibrator have been derived from
observations, and are used to remove the effects of the extraneous sources.
These models are updated periodically to allow for possible source
variability, and are used routinely in deriving the calibration
coefficients.

The most commonly used calibrators are listed in Table~5.  The flux
densities used for these sources are internally consistent and referenced
to 3C\,286. The flux densities adopted for the calibrators, based on
measurements made in 1994, are consistent with the scale of
Ott et al. (1994). The same sources are used to determine polarization
calibration parameters. The stability of telescope gain and phase is such
that it is only necessary to observe a calibrator prior to and following
each 12-hour observation.  Variations on shorter time-scales are corrected
using self-calibration techniques.

\begin{table}
\caption[]{Calibration sources}
\label{tabl-5}
\begin{center}
\begin{tabular}{lllll}
%\hline
\bf Source & $S_{408}$ (Jy) & $S_{1420}$ (Jy) & $p_{1420}$ (\%) & $\theta_{1420}$ (\degr) \\
3C\,48      & 38.9      & 15.7       &    0.6          & -- \\
3C\,147     & 48.0      & 22.0       &   $<0.25$       & -- \\
3C\,286     & not used  & 14.7       &    9.25         & 33.5 \\
3C\,295     & 54.0      & 22.1       &   $<0.25$       & -- \\
\multicolumn{5}{l}{Note: quoted flux densities are based on measurements} \\
\multicolumn{5}{l}{~~~~~~ made in 1994} \\
\end{tabular}
\end{center}
\end{table}

Calibration of the spectrometer is difficult because the narrow channel
bandwidth leads to high noise levels in individual channels. Calibration is
achieved in a two-step process, by using calibration coefficients measured
in the continuum bands. At the point where IF signals enter the central
telescope building, a broadband noise signal can be injected into all IF
paths, at a level sufficient to produce a correlation coefficient of
$\sim 0.3$, adequate to give an accurate measurement of gain and phase in
each spectrometer channel in 15 minutes. The continuum calibration measures
antenna-based gain and phase which vary relatively slowly with frequency,
but may vary with time. The IF calibration measures mostly the
channel-to-channel amplitude and phase differences, which arise largely in
the filters which define the bandpass, just before the signals are
digitized; these are quite stable with time. This IF calibration is
performed once every 4 days.

In later stages of image processing, amplitude and phase cor\-rections are
determined for individual antennas from self-calibration of Stokes $I$
images. These corrections are also applied to polarimeter images.  Under
some circumstances it is beneficial to apply them to \ion{H}{i} images
({\it e.g.\/} when there is a strong continuum source in the field).

\subsection{Antenna tracking and pointing}

The antennas are capable of tracking down to an elevation of 12\degr, so
the observable sky from the observatory's terrestrial latitude of 49\degr\
nominally extends from declination $-29$\degr\ to +89\degr\, (the latter
imposed by the telescope drive system).  However, full hour-angle coverage
is possible only for declinations north of +18\degr.  Sources north of
declination +54\degr\ are circumpolar, and there is an overlap in
hour-angle coverage at lower culmination, making it possible to track a
circumpolar source for 26 sidereal hours without interruption.
Non-sidereal tracking rates are available for observing solar-system
objects.

Since they are equatorially mounted, the antennas track only in hour angle,
although pointing corrections are made in both hour angle and declination.
A pointing model for each antenna accounts for zero offset and ellipticity
of the position encoders, misalignment of the polar axis, and
non-orthogonality of the declination and hour-angle axes.  These parameters
are empirically determined from ``nodding'' observations of selected
calibrators at many hour angles and declinations.  A ``nod'' consists of
measurements on-source and at 4 off-source positions situated 1\degr\
north, south, east, and west of the source, to which two-dimensional
gaussians are fitted to obtain errors in hour angle and declination.  The
r.m.s.\ pointing accuracy across the observable sky is $2.4'$ in R.A.\ and
$3.1'$ in Dec.

For a given observation, the pointing corrections at the declination of
interest are calculated from the pointing model for every 30 minutes of
hour angle, and interpolated linearly to intermediate hour angles.  The
tracking accuracy is then determined by a feedback loop in the control
system, which measures the deviation of the antenna position from the
requested position; at present errors in antenna position exceeding $3'$
are corrected.  Such corrections typically occur on timescales of 30
minutes.

\section{Performance}

\subsection{Synthesized beams}

The synthesized beam of the Synthesis Telescope depends on the taper
applied in the {\it u-v} plane; details of synthesized beams are given in
Table~6.  A uniform weighting is always applied when making images from the
telescope. The objectives of research with this telescope usually benefit
more from better sensitivity and lower sidelobe levels than from high
spatial resolution, so a Gaussian taper falling to 20\% at $144L$ is
typically applied to the {\it u-v} plane.

\begin{table*}
\caption[]{Details of synthesized beams}
\label{tabl-6}
\begin{center}
\begin{tabular}{lcccccc}
%Taper     & Frequency & Beamwidth & First & First & Second & Second \\
%          & ~~(MHz)   & (half power) & sidelobe & sidelobe & sidelobe & sidelobe \\
%          &           &              & level & radius & level & radius \\
Weighting & Frequency & Beamwidth & \multicolumn{2}{c}{First sidelobe} &
\multicolumn{2}{c}{Second sidelobe} \\

          & ~~(MHz)   & (half power) & level & radius & level & radius \\
   & & & & & & \\
Untapered    & 1420      & $49''$       & $-13$\% & $1'$   & 5.9\% & $1.65'$ \\
Untapered    & 408       & $2.8'$       & $-13$\% & $3.5'$ & 5.9\% & $5.85'$ \\
Gaussian$^*$ & 1420      & $58''$       & $-2$\%  & $1'$   & 2.4\% & $1.65'$ \\
Gaussian$^*$ & 408       & $3.4'$       & $-2$\%  & $3.5'$ & 2.4\% & $5.85'$ \\
 & & & & & & \\
\multicolumn{7}{l}{$^*$ A Gaussian taper falling to 20\% at $144L$
(617.1 m)} \\
\end{tabular}
\end{center}
\end{table*}

\subsection{Field of view}

The half-power beamwidth of the antennas is $107.2'$ at 1420\,MHz and
$332.1'$ at 408\,MHz, but ex\-per\-ience has shown that the usable field of
view extends at least to the 10\% level of the respective primary beams
(diameters of $187'$ and $578'$, respectively). Empirical measurements show
that these beams are well approximated by a function of the form $\cos^6
(q_\nu r)$. If $r$ is the radius in degrees, $q_{1420}=30.24$, and
$q_{408}=9.762$. Precise knowledge of the beam function is required for
accurate mosaicing of images of individual fields.

\subsection{Bandwidth and time-averaging effects}

Wide-field imaging is an important aspect of the Synthesis Telescope, so
off-axis effects must be carefully considered.

The continuum bands at 1420\,MHz span 35\,MHz, and, owing to differential
delay effects, if the entire band were assumed to be at the nominal centre
frequency there would be a reduction in point-source sensitivity of
approximately 50\% at a radius of $90'$.  This would be accompanied by a
radial smearing of a point source to approximately twice the nominal
beamwidth; while this preserves total flux, it is clearly unacceptable for
high-fidelity imaging.

By treating the four 7.5\,MHz continuum bands separately during imaging,
taking into account the actual centre frequency of each, the reduction in
point source sensitivity at $90'$ is only 5\%, with a comparable amount of
source distortion (see \cite{Bridle89}).  The 3.5-MHz bandwidth at 408\,MHz
reduces the point-source sensitivity by about 10\% at a radius of $300'$,
again with a similar amount of source distortion.

Effects due to time averaging are also a factor.  At the radii of $90'$ at
1420\,MHz and $300'$ at 408\,MHz, the visibility averaging period of 90\,s
reduces point-source sensitivity by about 8\% in a 12-hour observation,
with source distortion of similar magnitude in the azimuthal direction.
The combined effect of bandwidth and time-averaging smearing is then a
worst-case reduction of point-source sensitivity of about 13\% at 1420\,MHz
and 17\% at 408\,MHz, with both radial and azimuthal distortions.  Software
written for analyzing Synthesis Telescope images takes these effects into
account when determining, for example, point-source fluxes.

\subsection{Noise in Images}

The r.m.s.\ noise, $\Delta S$ (W\,m$^{-2}$\,Hz$^{-1}$), in an image made by
a synthesis telescope is (\cite{Crane89})
$$\Delta S = {{W\sqrt{2} k T_S}\over{\eta_c \eta_A A \sqrt{N_b N_{\rm IF} 
{\Delta f} \tau}}} \eqno(1)$$
where $W$ is a factor that depends on the weighting scheme applied to the
visibilities during imaging, $k$ is Boltzmann's constant, $T_S$ is the
system temperature (K), $\eta_c$ is the correlator efficiency (defined in
\S5), $\eta_A$ is the aperture efficiency of the antennas, each with area
$A$ (m$^2$), $N_b$ is the number of baselines, $N_{\rm IF}$ is the
number of IF channels, each of bandwidth $\Delta f$ (Hz), and $\tau$ is
the integration time (s). $N_{\rm IF}=1$ for a single polarization, and
$N_{\rm IF}=2$ when two polarizations are received.

The brightness-temperature sensitivity in K is
$$\Delta T = {{1}\over{\Omega}} {{\Delta S \lambda^2}\over{2k}}
\eqno(2)$$ 
where $\Omega$ is the synthesized beam area and $\lambda$ is the
wavelength.

The factor $W$ is unity for a naturally weighted image, but this weighting
scheme is never used with this telescope. Two weighting schemes used are
(a) uniform weighting, for which $W=2$, and (b) Gaussian tapering of
uniformly weighted data, tapering to 20\% at the longest spacing, for which
$W=1.33$.  The brightness temperature sensitivity is $\Delta
T=0.25\Delta S \sin\delta$\,K for case (a) and $\Delta
T=0.18\Delta S \sin\delta$\,K for case (b).

Table~7 shows the calculated sensitivities for each of the three telescope
outputs, and compares these predictions with measured noise levels from
images. At 408\,MHz the thermal noise limit is not reached because of
confusion.

\begin{table*}
\caption[]{Telescope sensitivity}
\label{tabl-7}
\begin{center}
\begin{tabular}{lccc}
  & 1420\,MHz       & 1420\,MHz           & 408\,MHz      \\
  & continuum       & spectrometer        & continuum    \\
  &                 &                     &              \\
$T_S$, K            & 60    & 60          & 150$^*$   \\
$\eta_A$            & 0.55  & 0.55        & 0.60      \\
$\eta_c$            & 0.985 & 0.88        & 0.88      \\
$N_b$               & 21    & 12          & 21        \\
$N_{IF}$            & 2     & 2           & 1         \\
$\Delta f$, MHz     & 30    & $B/160^\dagger$ & 3.5 \\
$W$                 & 2.0   & 1.33        & 2.0       \\
$\Delta S_{theor}$ mJy/beam & 0.28 & $20B^{-0.5}$ & 3.0 \\
$\Delta T_{theor}$ K & $0.071\sin\delta$ & $3.5B^{-0.5}\sin\delta$ & $0.75\sin\delta$ \\
$\Delta S_{meas}$ mJy/beam  & 0.27 & $18B^{-0.5}$  & 3.8 \\
%$\Delta T_{meas}$ K         &      &             &           \\
 & & & \\
\multicolumn{4}{l}{$^*$ Includes $T_{sky}=45$\,K, a typical value in the Galactic plane}  \\
\multicolumn{3}{l}{$^\dagger$ $B$ is the overall spectrometer bandwidth 
in MHz} & \\

\end{tabular}
\end{center}
\end{table*}

The dynamic range achieved with the telescope varies with frequency and
region observed. In the 1420\,MHz continuum channel, dynamic ranges of
$10^4$ are commonly achieved. The limits to this performance are, at the
low end, system noise, and, at the high end, artefacts in the vicinity of
strong sources. Artefacts, which are very difficult to remove by normal
processing, rise above the thermal noise limit when telescope pointing is
closer than 3\degr\ to Cas A or Cyg A. At 408 MHz, the dynamic range
achieved is about $5 \times 10^3$, for offsets greater than 5\degr\ from
strong sources.

\subsection{Imaging Performance}

Owing to the thorough sampling of baselines in a $12\times12$~hour
observation set, the images produced by the Synthesis Telescope have very
low sidelobe levels and excellent sensitivity to extended structure. This
can be seen clearly in Figure~7, which shows part of the data from one
1420-MHz field from the Canadian Galactic Plane Survey. The extended object
which dominates Stokes $I$ images of this field is the \ion{H}{ii} region
W5 (IC~1848), in the Perseus arm of the Galaxy at a distance of
$\sim2$\,kpc.  The ``raw'' image (top left), is made from the calibrated
visibilities after editing interference, but before application of any
image processing routines.  The large emission regions are surrounded by
depressions, produced by the lack of data for baselines $< 3L$. The final
image, after incorporation of single-antenna data, is shown at upper right.

The raw Stokes $U$ image (lower left) shows structure on many scales,
totally unrelated to that seen in the $I$ image. These polarization
structures arise from Faraday rotation along the line of sight -- they are
not intrinsic features of the emission source.  The large elliptical
polarization structure is discussed briefly in \S10.

The lower right panel shows one of the 256 spectral-line channels; the data
have been edited and a continuum image, formed from spectrometer channels
free of \ion{H}{i} emission, has been subtracted, but no single-antenna
data have been added.

\begin{figure*}
\resizebox{16cm}{!}{\includegraphics{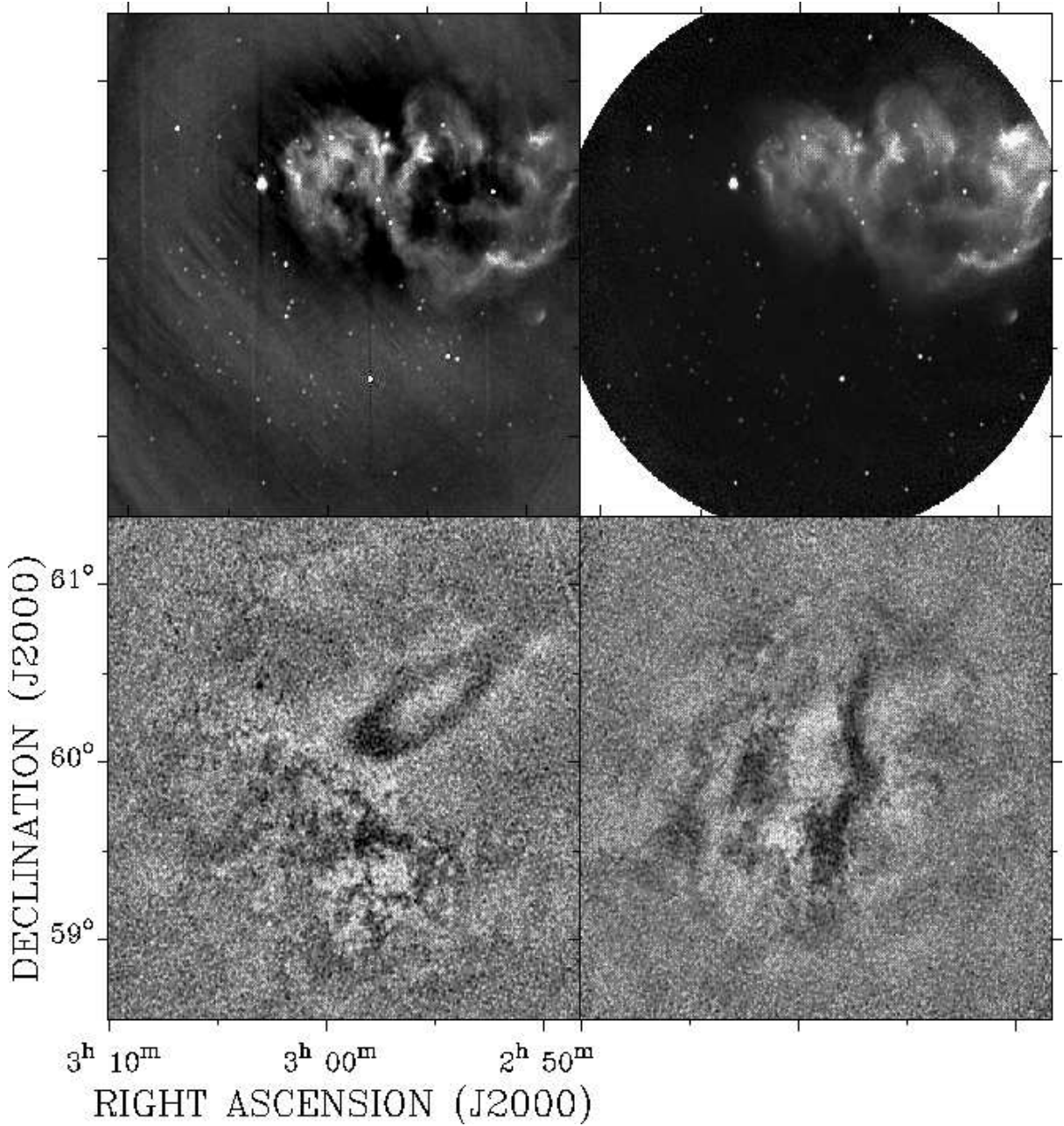}}
\caption{Images at 1420\,MHz of a field containing the \ion{H}{ii} region
W5.  Top left: continuum image made from calibrated visibilities after
editing interference, but before application of any image processing
routines. Top right: fully processed image with single-antenna data
incorporated. A correction has been applied to compensate for the primary
beam of the antennas. The circular boundary is at a radius of $90'$. Lower
left: image in Stokes parameter $U$.  Other than editing of interference,
no processing has been applied. Lower right: \ion{H}{i} image in one of 256
channels of the spectrometer. Continuum emission has been estimated from
end-channels of the spectrometer and subtracted. No single-antenna data
have been added to either of the lower two images.}
\label{fig-7}
\end{figure*}

\section{Astronomical applications}

We have described a synthesis telescope uniquely suited to observation of
the Galactic ISM.  While its resolution ($\sim1'$) is modest compared to
some synthesis telescopes ({\it e.g.\/} the VLA), its sensitivity to
extended structure is superior, and it offers significantly different
performance from single-antenna telescopes.  In this section we discuss a
few examples of results obtained with the telescope, chosen to illustrate
its unique capabilities.  A more detailed discussion will be found in
Taylor et~al.\ (2000).

At distances of 2 to 10\,kpc, the $1'$ angular resolution of the
telescope translates into a physical resolution of 0.6 to 3\,pc, which
has proved to be a good match to significant ISM structures.  The
interstellar \ion{H}{i} shows extensive filamentary structure on these
scales both within the Galactic plane (\cite{Normand97}) and at high
latitudes (\cite{Joncas92}). While the origin of the fine structure in
this emission is not completely understood, it seems very often to be
related to interfaces between the atomic phase and some other
constituent of the ISM, or to energy injection, as from stellar winds.

Experience with the telescope has shown that there are significant ISM
phenomena which extend over many degrees, but require arcminute
angular resolution for their detection.  An example is provided by the
discovery of a Galactic chimney by Normandeau et~al.\ (1996). This
structure is a conduit for radiation and material from the disk to the
halo, and it is outlined by a sharp interface between the \ion{H}{i}
exterior and the ionized interior.  Its size is $\sim5$\degr,
equivalent to several hundred pc at the distance of the Perseus Arm
($\sim2$\,kpc) .  It is nevertheless extremely difficult to detect in
the survey data of Hartmann \& Burton (1997): even though that survey
has much higher nominal sensitivity than the Synthesis Telescope, its
angular resolution is only $36'$.

Similar considerations apply to the detection of thermal and non-thermal
continuum objects.  The continuum sensitivity of the telescope (listed in
Table~7) translates into sensitivity to thermal emission measure (at both
frequencies) of $\sim27$\,cm$^{-6}$\,pc and sensitivity to synchrotron
emission of $\sim4 \times 10^{-23}$\,W\,m$^{-2}$\,Hz$^{-1}$\,sr$^{-1}$,
also at both frequencies.  An example is provided by the detection of large
supernova remnants (SNRs) of low surface brightness.  For example,
\cite{Landecker90} discovered the SNR G65.1+0.6, whose size is $90' \times
51'$.  The overall surface brightness of this object is about half of the
nominal rms sensitivity quoted above, indicating the importance of
filamentary structure in making such objects detectable.

Polarization imaging at 1420\,MHz allows detailed analyses of discrete
emitters of polarized radiation, particularly SNRs.  For example, \cite{Leahy97}
have mapped the polarized emission from the full extent of the Cygnus Loop.
However, only part of the SNR is strongly polarized, with some of the shell
emission depolarized by strong Faraday rotation within a thermal electron
component mixed in with the compressed fields and synchrotron emitting
particles.

While SNRs are prominent objects in Stokes $I$ images, most are barely
significant in images of Stokes $Q$ or $U$.  The dominant feature in
all polarization images made close to the Galactic Plane is
widespread, low-level structure, seen most prominently in polarization
angle.  This structure is understood as the result of Faraday rotation
in the magnetized ISM acting on polarized signals from the diffuse
Galactic synchrotron radiation.  It tells us less about the emitter
than it does about the medium through which the polarized radiation
has travelled.

The study of this phenomenon is providing a new window on the ISM,
through its sensitivity to magnetic fields and low-density ionized
material.  The frequency of 1420 MHz seems ideal for such studies
near the Galactic plane, where the properties of the Faraday screen
lead to a change in polarization angle that is neither too
small to produce measurable changes nor so large that it causes
depolarization within the beam or the bandwidth of the telescope.  The
large elliptical feature seen in Figure~7 (lower left panel), for
example, which is very uniform in structure in contrast to its chaotic
surroundings, is understood as just such a manifestation of ionized
material in the inter-arm region (\cite{Gray98}).  Other polarization
results (\cite{Gray99}) have led to the detection of an extended
envelope of a large \ion{H}{ii} region, and the measurement of the
magnetic field in that envelope. In both cases, the \ion{H}{ii}
regions themselves produce depolarization due to the turbulent,
high-density ionized material within them.

\section*{\it Acknowledgements} 

The development and construction of the DRAO Synthesis Telescope has
been the work of many people over many years. Notable for their skill
and dedication have been Jean Bastien, Ron Casorso, Ed Danallanko,
Jack Dawson, Ron McDougall, Harry Mielke, Diane Parchomchuk, Ev
Sheehan and Rod Stuart. The work has also involved many students and
others who have worked at DRAO for shorter periods. We are indebted to
them all.  The DRAO Synthesis Telescope is operated as a national
facility by the National Research Council of Canada. Graduate students
who have worked on the development of the telescope have been
supported by grants to TLL, DR, and JFV from the Natural Science and
Engineering Research Council.

\end{document}